\documentclass[preprint,showpacs,preprintnumbers,amsmath,amssymb,showkeys]{revtex4}

\usepackage{graphicx}
\usepackage{dcolumn}
\usepackage{bm}

\begin{document}

\preprint{}

\title{Polarization-sensitive propagation in an anisotropic metamaterial with double-sheeted hyperboloid dispersion relation}
\author{Hailu Luo}\email{
E-mail: hailuluo@gmail.com (H. Luo)}
\author{Zhongzhou Ren}
\affiliation{ Department of Physics, Nanjing University, Nanjing
210008, China}
\date{\today}

\begin{abstract}
The polarization-sensitive propagation in the anisotropic
metamaterial (AMM) with double-sheeted hyperboloid dispersion
relation is investigated from a purely wave propagation point of
view. We show that TE and TM polarized waves present significantly
different characteristics which depend on the polarization.  The
omnidirectional total reflection and oblique total transmission can
occur in the interface associated with the AMM. If appropriate
conditions are satisfied, one polarized wave exhibits the total
refraction, while the other presents the total reflection. We find
that the opposite amphoteric refractions can be realized by rotating
the principle axis of AMM, such that one polarized wave performs the
negative refraction, while the other undergoes positive refraction.
The polarization-sensitive characteristics allow us to construct two
types of efficient polarizing beam splitters under certain
achievable conditions.
\end{abstract}

\maketitle

\section{Introduction}\label{Introduction}
Media with negative permittivity and permeability have not been
found in violation of any fundamental physical principles have been
verified both experimentally and
theoretically~\cite{Veselago1968,Smith2000,Shelby2001,Parazzoli2003,Houck2003}.
In such materials, the directions of energy transfer and wavefront
propagation are opposite. This leads to remarkable electromagnetic
properties such as refraction at surfaces that is described by a
negative refraction index. It was also found negative refraction can
occur at an interface associated with an anisotropic metamaterial
(AMM), which does not necessarily require that all tensor elements
of $\boldsymbol{\varepsilon}$ and $\boldsymbol{\mu}$ have negative
values~\cite{Lindell2001,Hu2002,Smith2003,Thomas2005}. In general,
TE and TM polarized waves propagate in different directions in an
anisotropic medium. For a conventional anisotropic medium, all
tensor elements of permittivity $\boldsymbol{\varepsilon}$ and
permeability $\boldsymbol{\mu}$ are positive. Wave propagation in
conventual anisotropic crystal is an interesting topic with both a
conceptual and a practical value. A large number of optical devices
are based on the anisotropic effect, such as polarizers,
compensators, switches etc., are currently employed in a large
amount of experimental situations~\cite{Yariv1984}. Recently, a
broad range of applications have been suggested, such as partial
focus lens~\cite{Smith2004,Parazzoli2004,Dumelow2005,Luo2007a},
spatial filters~\cite{Schurig2003,Martinez2005}, and polarizing beam
splitters~\cite{Hu2006,Luo2007b} can be realized by AMMs.

In classic electrodynamics, it is well known that the three
dimensional (3D) frequency contour of anisotropic materials is the
combination of sphere and ellipsoid~\cite{Chen1983,Landau1984}.
Since the advent of negative media parameters in AMMs, the 3D
frequency contour are significantly different from conventional
anisotropic media. The corresponding wave-vector surfaces are a
combination of ellipsoid or single-sheeted hyperboloid or
double-sheeted hyperboloid. In the development of AMMs, there are
several important questions easily be inquired: how the waves
behave in AMMs, what characteristics may be useful for practical
applications, and how to construct such AMMs. The importance of
investigating AMMs with new wave-vector surfaces and its potential
applications becomes evident when one considers that low-loss
optical metamaterials are increasingly
possible~\cite{Doling2007,Chettiar2007,Soukoulis2007,Lezec2007}.
Here we will focuss our attention on the case that both TE and TM
polarized waves exhibit double-sheeted hyperboloid wave-vector
surfaces.

In this work, we want to present an investigation on the
polarization-sensitive characteristics in the AMM with
double-sheeted hyperboloid dispersion relation. First, we want to
explore the wave propagation in the AMM with different combinations
of tensor elements. We show that the omnidirectional total
reflection and oblique total transmission can occur in the interface
associated with AMM. If certain conditions are satisfied, one
polarized wave exhibits the total refraction, while the other
presents the total reflection. Next, we concentrate our interest on
the amphoteric refractions. We find that the opposite amphoteric
refractions can be realized by rotating the principle axis of AMM,
such that one polarized wave performs negative refraction, while the
other undergoes positive refraction. Finally, we study the practical
applications of the polarization-sensitive characteristics, and two
types of polarizing beam splitters can be constructed.

\section{Wave propagation in anisotropic metamaterials}\label{sec2}
To reveal the phenomenon of polarization-sensitive propagation, we
start with a purely 3D wave propagation point of view. It is
currently well accepted that a better model is to consider
anisotropic constitutive parameters, which can be diagonalized in
the coordinate system collinear with the principal axis of the
metamaterial~\cite{Smith2003,Thomas2005}. If we take the principal
axis as $z$ axis, the permittivity and permeability tensors have
the following forms:
\begin{eqnarray}
\boldsymbol{\varepsilon}=\left(
\begin{array}{ccc}
\varepsilon_x  &0 &0 \\
0 & \varepsilon_y &0\\
0 &0 & \varepsilon_z
\end{array}
\right), ~~~\boldsymbol{\mu}=\left(
\begin{array}{ccc}
\mu_x &0 &0 \\
0 & \mu_y &0\\
0 &0 & \mu_z
\end{array}
\right),\label{matrix}
\end{eqnarray}
where $\varepsilon_i$ and $\mu_i$ are the permittivity and
permeability constants in the principal coordinate system
($i=x,y,z$).  In general, it is not enough for a complex
metamaterial be characterized by six tensor elements, since the
response of anisotropic metamaterial can be very complex. But for a
special case, however, it is enough. Since we restrict the wave
propagation at the $x-z$ plane. For a certain polarized wave, the
propagation only decided by certain three parameters, other three
parameters do not intervene.

Consider the propagation of a planar wave of frequency $\omega$ as
${\bf E}={\bf E}_0 e^{i {\bf k} \cdot {\bf r}-i \omega t}$ and ${\bf
H} = {\bf H}_0 e^{i {\bf k} \cdot {\bf r}-i\omega t}$, through a
regular isotropic medium toward an AMM. In isotropic media, the
accompanying dispersion relation has the familiar form
\begin{equation}
 k_{x}^2+ k_{y}^2+k_{z}^2={\varepsilon_I
\mu_I} \frac{\omega^2}{c^2}. \label{D1}
\end{equation}
Here $k_i$ is the $i$ component of the incident wave vector, $c$ is
the speed of light in vacuum, $\varepsilon_I$ and $\mu_I$ are the
permittivity and permeability, respectively. Electromagnetic waves
can be classified into two types: TE and TM waves. For a TE wave,
its electric field is perpendicular to the plane of incidence. For a
TM wave, its magnetic field is normal to the plane of incidence. In
isotropic media, both TE and TM waves exhibit the same dispersion
relation. A careful calculation of Maxwell's equations in
anisotropic media gives the dispersion relations:
\begin{eqnarray}
\frac{ q_{x}^2}{\varepsilon_y \mu_z}+\frac{q_{y}^2}{\varepsilon_x
\mu_z}+\frac{q_{z}^2}{\varepsilon_y
\mu_x}-\frac{\omega^2}{c^2}=0,\label{DE2}\\
\frac{q_{x}^2}{\varepsilon_z \mu_y}+\frac{q_{y}^2}{\varepsilon_z
\mu_x}+\frac{q_{z}^2}{\varepsilon_x
\mu_y}-\frac{\omega^2}{c^2}=0,\label{DH2}
\end{eqnarray}
for TE and TM polarized waves, respectively~\cite{Luo2007c}. Here
$q_{i}$ represents the $i$ component of transmitted wave-vector.
The above equations can be represented by two three-dimensional
surfaces in wave-vector space. It is well known that the wave
propagation behaviors in anisotropic media are significantly
different from those in isotropic
materials~\cite{Chen1983,Landau1984}. In general, anisotropic
media exhibit a polarization-sensitive effect, namely TE and TM
polarized waves present different characteristics which depend on
the polarization. Here we are particularly interested in the case
that one polarized wave exhibit total refraction, while the other
present total reflection. Furthermore, we will discuss that the
two polarized waves experience an opposite amphoteric refraction,
such as polarized wave performs negative refraction, while the
other undergoes positive refraction.

First we explore the two polarized waves show anomalous total
refraction or total reflection. Without loss of generality, we
assume the wave vector locate at the $x-z$  plane ($k_y=q_y=0$). The
incidence angle of light is given by $\theta_I
=\tan^{-1}[{k_x}/{k_{z}}]$. Based on the boundary condition, the
tangential components of the wave vectors must be continuous, i.e.,
$q_x=k_x$. Then the refraction angle of the transmitted wave vector
or phase of TE and TM polarized waves can be written as
$\beta_P^{TE}=\tan^{-1}[{q_x}/{q_{z}^{TE}}]$ and
$\beta_P^{TM}=\tan^{-1}[{q_x}/{q_{z}^{TM}}]$, respectively. It
should be noted that the actual direction of light is defined by the
time-averaged Poynting vector ${\bf S} =\frac{1}{2} {Re}({\bf
E}\times \bf{H}^\ast)$~\cite{Jackson1998}.  The refraction angle of
Poynting vector can be obtained as $\beta_S^{TE}=
\tan^{-1}[{S_{Tx}^{TE}}/{S_{Tz}^{TE}}]$ and $\beta_S^{TM}=
\tan^{-1}[{S_{Tx}^{TM}}/{S_{Tz}^{TM}}]$ for TE and TM waves,
respectively.

In principle, the occurrence of transmission requires that the $z$
component of the wave vector must be real. Setting $q_z=0$ in
Eq.~(\ref{DE2}) and Eq.~(\ref{DH2}), we can obtain the critical
angles as
\begin{equation}
\theta_C^{TE}=\sin^{-1}\left[\sqrt{\frac{\varepsilon_y
\mu_z}{\varepsilon_I
\mu_I}}\right],~~~\theta_C^{TM}=\sin^{-1}\left[\sqrt{\frac{\varepsilon_z
\mu_y}{\varepsilon_I \mu_I}}\right].\label{CA}
\end{equation}
If numerator larger than denominator inside the square root, no wave
can transmitted for any incident angle, and the AMM will exhibit the
interesting phenomenon of omnidirectional total reflection. Here the
omnidirectional total reflection means that the total reflection
occurs for wave incidents at any angles. In addition, the oblique
total reflection can not exist for a certain material parameters
which make the corresponding expressions inside the square root
negative.

The transmitted wave in AMMs can be determined by the two
principles~\cite{Born1999}: First, the boundary conditions require
that the tangential component of the wave vector, is conserved
across the interface, $q_x=k_x$. Second, energy conservation law
requires that the energy current of the refracted waves should
transmit away from the interface, i.e., the normal component of
the Poynting vector, $S_{Tz}>0$. For TE olarized wave, the
transmitted Poynting vector is given by
\begin{equation}
{\bf S}_T^{TE}=Re\left[\frac{T_{TE}^2 E_0^2 q_x^{TE}}{2 \omega
\mu_z}{\bf e}_x+\frac{T_{TE}^2 E_0^2 q_z^{TE}}{2\omega\mu_x}{\bf
e}_z\right].\label{SE}
\end{equation}
Analogously, for TM polarized wave, the transmitted Poynting vector
is given by
\begin{equation}
{\bf S}_T^{TM}=Re\left[\frac{T_{TM}^2 H_0^2 q_x^{TM}}{2
\omega\varepsilon_z}{\bf e}_x+\frac{T_{TM}^2 H_0^2
q_z^{TM}}{2\omega\varepsilon_x}{\bf e}_z\right].\label{SH}
\end{equation}
Here $T_{TE}$ and $T_{TM}$ are the transmission coefficient for TE
and TM polarized waves, respectively. Based on the boundary
condition, we can obtain the following expression for the
transmission coefficients:
\begin{equation}
T_{TE} = \frac{2 \mu_x k_z}{\mu_x  k_z+\mu_I q_z^{TE}},~~~~~
T_{TM} = \frac{2 \varepsilon_x k_z}{\varepsilon_x
k_z+\varepsilon_I q_z^{TM}}.\label{TETH}
\end{equation}
Simultaneously, the corresponding reflection coefficients can be
obtained as
\begin{equation}
R_{TE}=\frac{\mu_x k_z-\mu_I q_z^{TE}}{\mu_x  k_z+ \mu_I
q_z^{TE}},~~~ R_{TM}=\frac{\varepsilon_x k_z-\varepsilon_I
q_z^{TM}}{\varepsilon_x k_z+\varepsilon_I q_z^{TM}}.\label{RERH}
\end{equation}
It should be noted that the oblique total refraction emerges when a
wave incident at Brewster angle. Mathematically the Brewster angles
can be obtained from $R_{TE}=0$ and $R_{TM}=0$. In the next section,
we will pay our attention to the case that TE polarized wave
experiences oblique total refraction, while TM polarized wave
undergoes total reflection, i.e. $R_{TE}=0$. The anomalous wave
propagation depends on the choice of the anisotropic parameters. In
general, the critical angle is larger than the Brewster angle in
regular anisotropic media. In present AMM, however, the situation is
reversed. Here the anomalous wave propagation means that the
Brewster angle is larger than the critical angle.

Next we want to study the opposite amphoteric refractions. Unlike
in isotropic media, the Poynting vector in AMMs is neither
parallel nor antiparallel to the wave vector, but rather makes
either an acute or an obtuse angle with respect to the wave
vector~\cite{Lindell2001}. In general, to distinguish the positive
and negative refractions in AMMs, we must calculate the direction
of the Poynting vector with respect to the wave vector. Positive
refraction means ${\bf q}_x\cdot{\bf S}_{T}>0$, and anomalous
negative refraction suggests ${\bf q}_x\cdot{\bf S}_{T}<0$. From
Eqs.~(\ref{SE}) and (\ref{SH}) we get
\begin{equation}
{\bf q}_x^{TE}\cdot{\bf S}_{T}^{TE}=\frac{T_{TE}^2 E_0^2
(q_x^{TE})^2}{2 \omega \mu_z},~~~ {\bf q}_x^{TM}\cdot{\bf
S}_{T}^{TM}=\frac{T_{TM}^2 H_0^2 (q_x^{TM})^2}{2 \omega
\varepsilon_z}.
\end{equation}
We can see that the amphoteric refractions will be determined by
$\mu_z$ for TE polarized wave and $\varepsilon_z$ for TM polarized
wave. The underlying secret of the opposite amphoteric refractions
is that $\varepsilon_z$ and $\mu_z$ always have the opposite signs.
Evidently, we can choose an appropriate combinations of the tensor
elements to realize the opposite amphoteric refractions. While we
are particularly interested in the case that the amphoteric
refractions present in the AMM with rotating principle axis. In the
section IV, we will discuss such an interesting phenomenon in
details.

\section{Double-sheeted hyperboloid dispersion relation}\label{sec3}
In this section, we will explore the polarization-sensitive
propagation in three types of AMMs. It should be noted that, for
the same dispersion relation there exist two subtypes 3D
wave-vector surface which can be formed from combinations of
tensor elements. In fact, the two subtypes can be discussed in
similar way. Hence we do not wish to get involved in the trouble
to discuss every subtypes in detail.

\begin{figure}
\includegraphics[width=6cm]{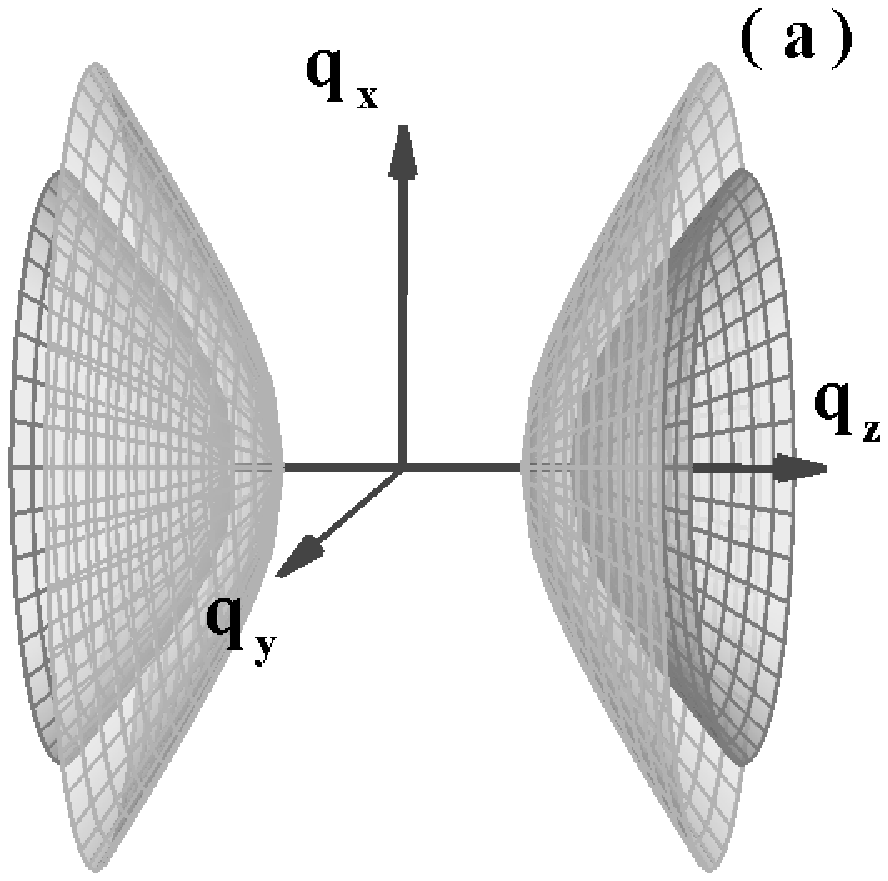}~~~~~~~~
\includegraphics[width=6cm]{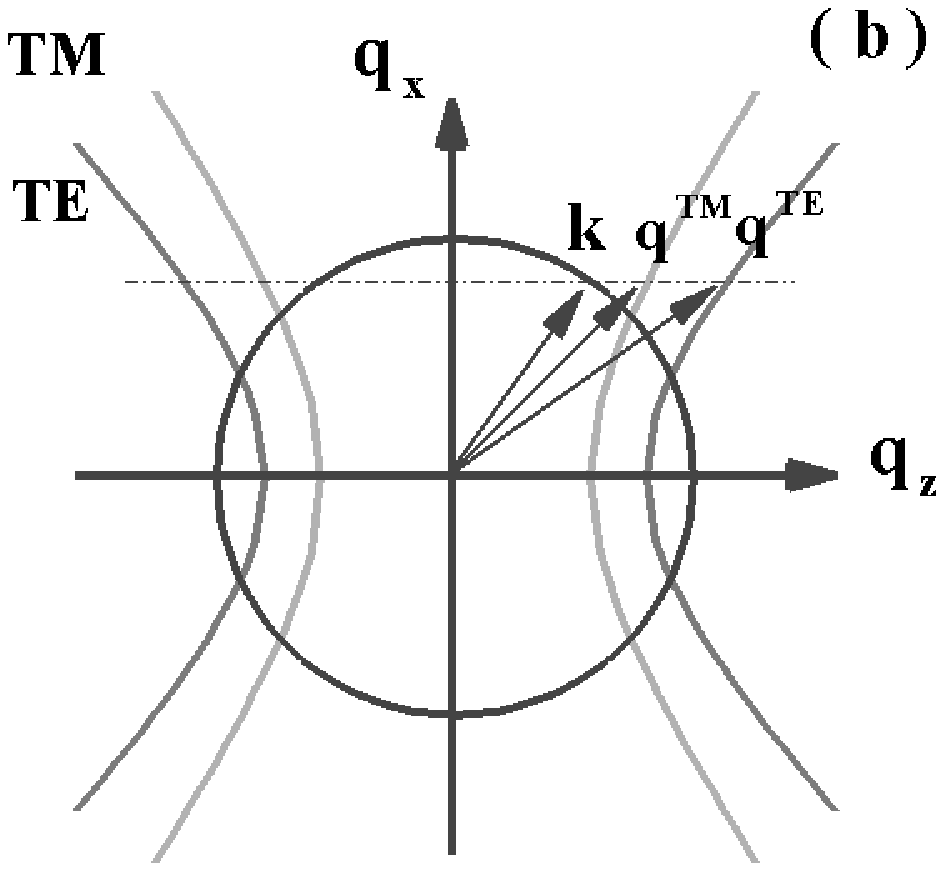}
\caption{\label{Fig1} The two double-sheeted hyperboloid have the
same revolution axis. We assume the revolution axis coincide with
$z$ axis. (a) The 3D frequency contours present propagation
characteristics of TE (black) and TM (gray) waves. (b) The circle
and the double-sheeted hyperbola represent the dispersion relations
of isotropic regular media and anisotropic metamaterial,
respectively.}
\end{figure}

Case I.~  The two double-sheeted hyperboloid have the same
revolution axis. Here we choose the revolution axis coincide with
$z$ axis. The corresponding combination is chosen as
$\boldsymbol{\varepsilon}=[+,+,-]$ and $\boldsymbol{\mu}=[+,+,-]$.
The 3D frequency contour is plotted in Fig.~\ref{Fig1}(a). To
investigated the propagating behaviors in this kind of AMM, we
plotted the 2D refraction diagram in Fig.~\ref{Fig1}(b). The
circle and the double-sheeted hyperbola represent the dispersion
relations of isotropic regular media and AMMs, respectively. Both
TE and TM polarized waves occur in the branch
$-\pi/2<\theta_{I}<\pi/2$.  For the two polarized waves, ${\bf
k}_z\cdot{\bf q}_{z}>0$ and ${\bf q}_x\cdot{\bf S}_{T}<0$, so the
wave-vectors exhibit positive refractions, whereas Poynting
vectors undergo negative refractions.

Case II. The two double-sheeted hyperboloid have the same revolution
axis. Here we choose the revolution axis coincide with $x$ axis. The
corresponding combination is given by
$\boldsymbol{\varepsilon}=[-,+,+]$ and $\boldsymbol{\mu}=[-,+,+]$.
The 3D frequency contour is plotted in Fig.~\ref{Fig1}(a). To
investigated the propagating behaviors, we depict the 2D refraction
diagram in Fig.~\ref{Fig1}(b). For the two polarized waves, ${\bf
k}_z\cdot{\bf q}_{z}<0$ and ${\bf q}_x\cdot{\bf S}_T>0$, so their
refractions of wave vectors are negative, while the refractions of
Poynting vectors are always positive. For TE polarized waves, if
$\varepsilon_z \mu_y<\varepsilon_I \mu_I$ the propagation occur in
the branch $-\pi/2<\theta_{I}<-\theta_{C}^{TE}$ and
$\theta_{C}^{TE}<\theta_{I}<\pi/2$. Note that the oblique total
refraction can occur in the branch
$-\theta_C^{TE}<\theta_I<\theta_C^{TE}$. If $\varepsilon_z
\mu_y>\varepsilon_I \mu_I$, no wave can transmitted for any incident
angle, and the AMM will exhibit the interesting phenomenon of
omnidirectional total reflection. For TM polarized waves, if the
inequality $\varepsilon_z \mu_y<\varepsilon_I \mu_I$, the
propagation occurs in the branch
$-\pi/2<\theta_{I}<-\theta_{C}^{TM}$ and
$\theta_{C}^{TM}<\theta_{I}<\pi/2$. If $\varepsilon_z
\mu_y<\varepsilon_I \mu_I<\varepsilon_y \mu_z$, when
$\theta_C^{TE}<\theta_I<\theta_C^{TM}$, only the TE polarized wave
can propagate into the AMM for a certain incidence branch, while the
TM polarized wave is totally reflected for any incidence angle.

\begin{figure}
\includegraphics[width=6cm]{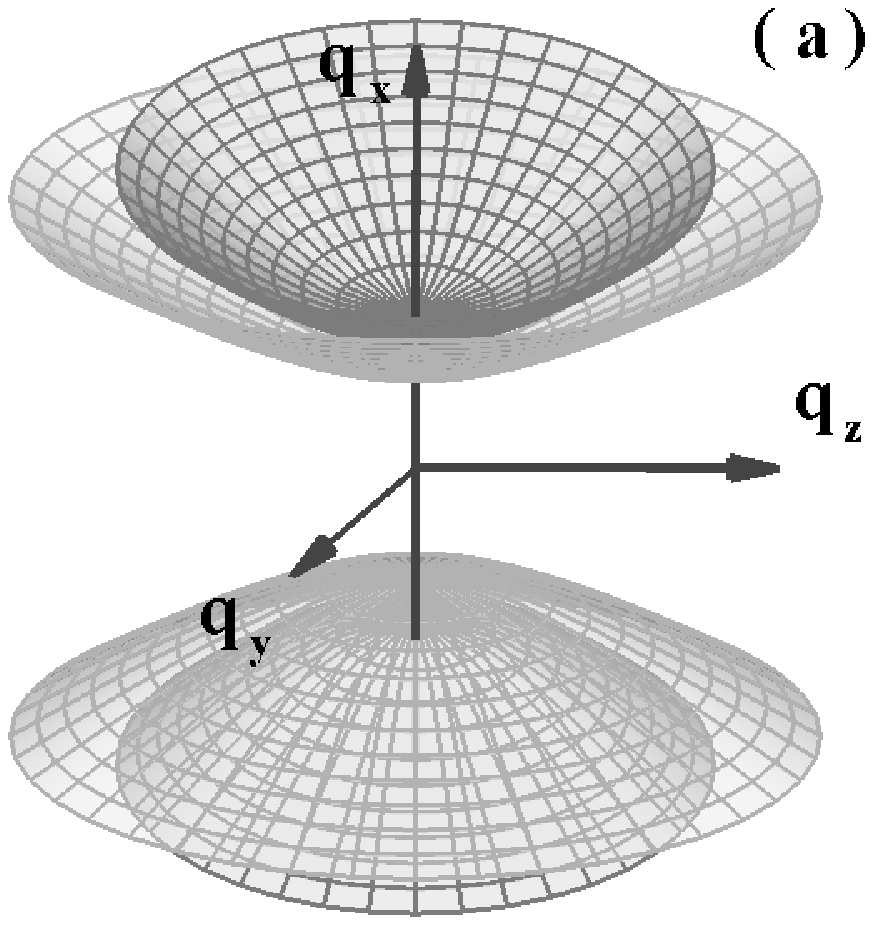}~~~~~~~~
\includegraphics[width=6cm]{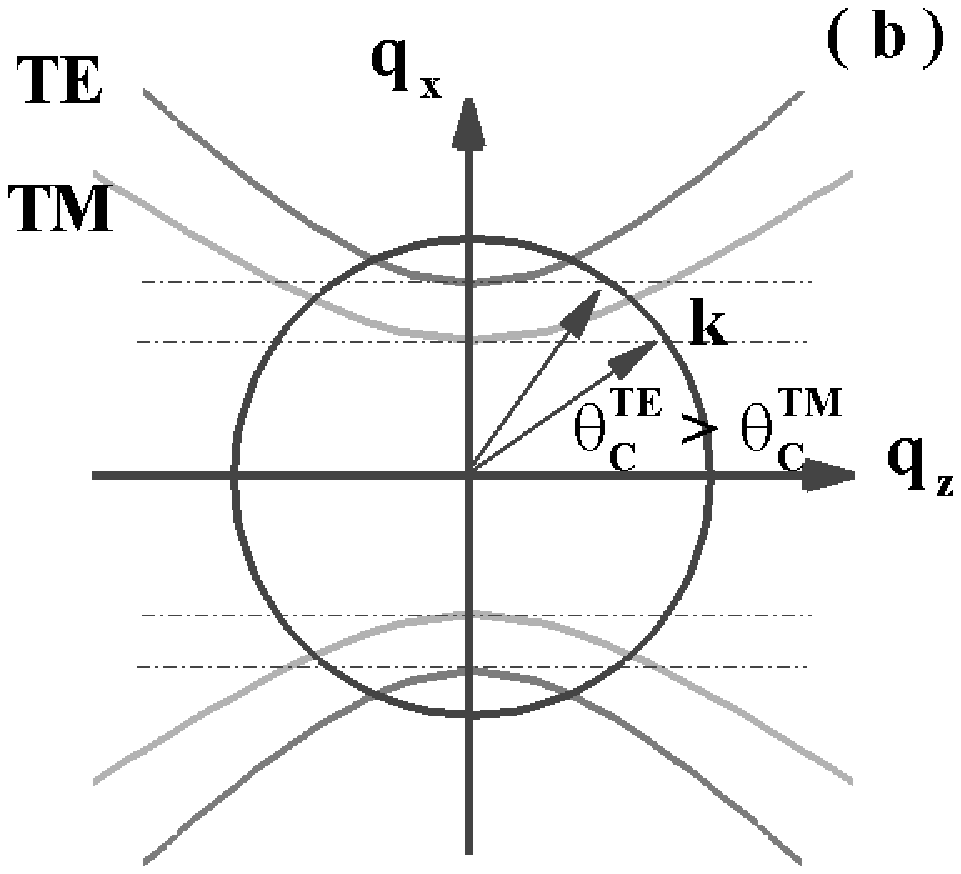}
\caption{\label{Fig2} The two double-sheeted hyperboloid have the
same revolution axis. TE and TM waves have the same revolution axis
which coincide with $x$ axis. (a) The 3D frequency contours present
propagation characteristics of TE (black) and TM (gray) waves. (b)
The circle and the double-sheeted hyperbolas represent the
dispersion relations of isotropic regular media and anisotropic
metamaterial, respectively. }
\end{figure}

\begin{figure}
\includegraphics[width=6cm]{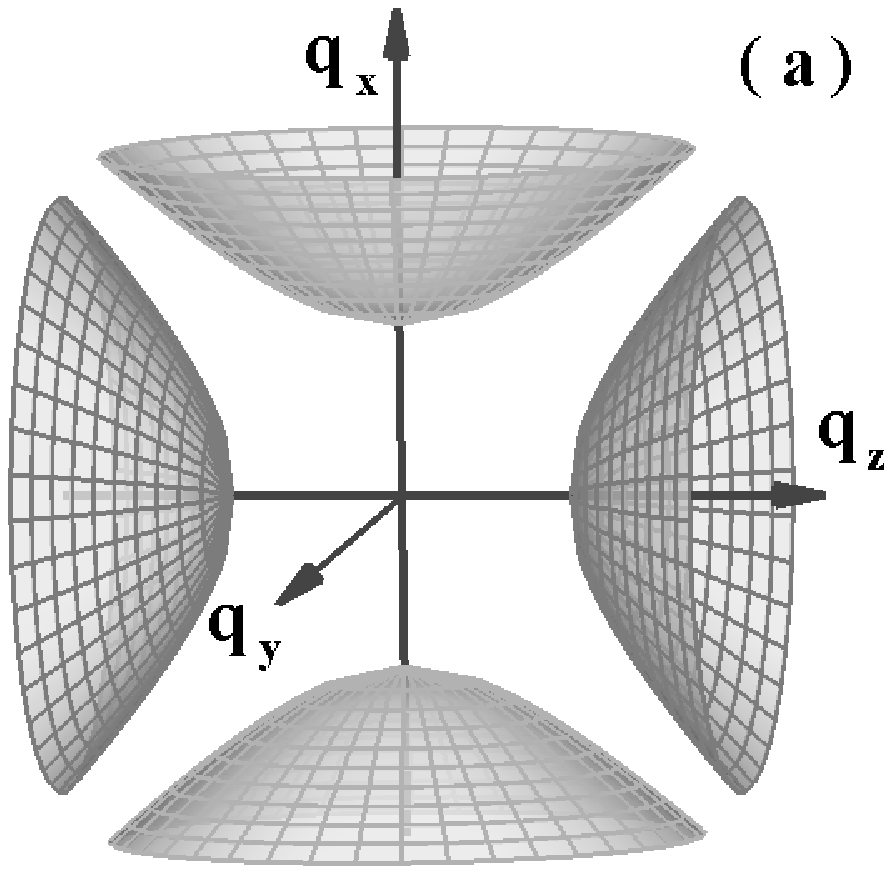}~~~~~~~~
\includegraphics[width=6cm]{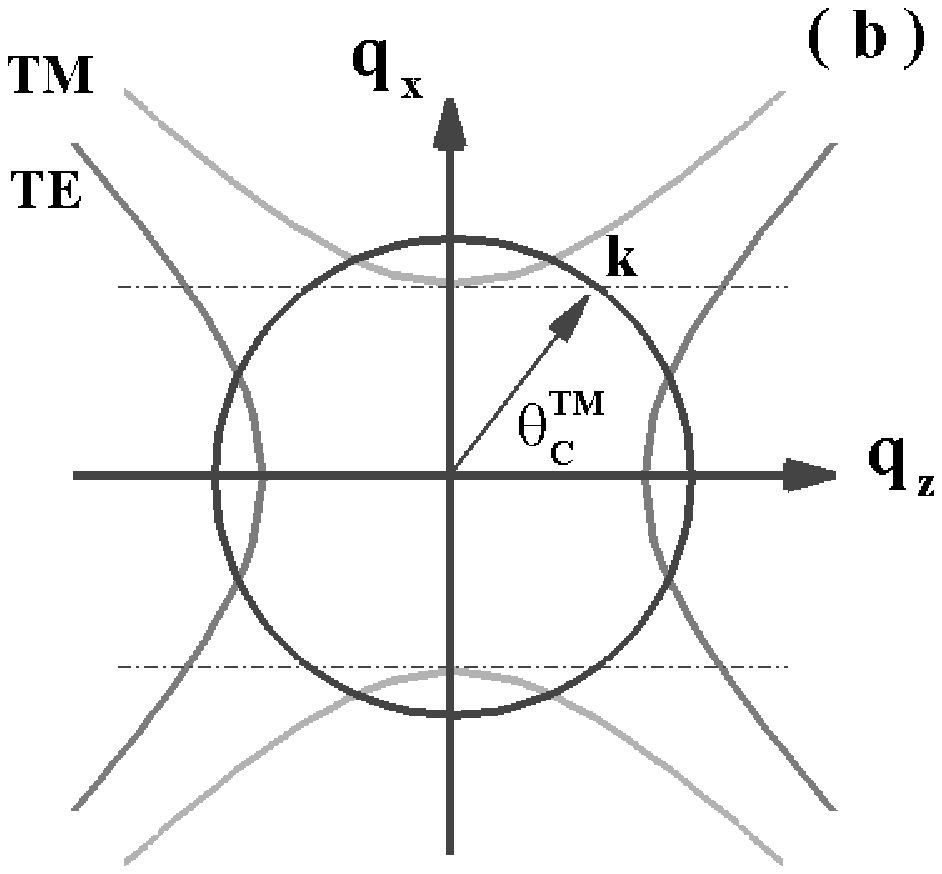}
\caption{\label{Fig3} The revolution axes of the two double-sheeted
hyperboloid are perpendicular to each other. The revolution axis of
TE and TM waves coincide with $x$ and $z$ axis, respectively. (a)
The 3D frequency contours present propagation characteristics of TE
(black) and TM (gray) waves. (b) The circle and the double-sheeted
hyperbolas represent the dispersion relations of isotropic regular
media and anisotropic metamaterial, respectively.}
\end{figure}

Case III. The revolution axes of the two double-sheeted hyperboloid
are perpendicular to each other. As an example, we want to explore
the combination with $\boldsymbol{\varepsilon}=[+,+,-]$ and
$\boldsymbol{\mu}=[+,-,-]$. The 3D frequency contour is plotted in
Fig.~\ref{Fig3}(a). For TE polarized wave, $\varepsilon_y \mu_z$ is
negative, thus the inequality $\varepsilon_y \mu_z<\varepsilon_I
\mu_I$ satisfied for any incidence angle. In this case, the real
wave vector exists for the branch $-\pi/2<\theta_{I}<\pi/2$. Here
${\bf k}_z\cdot{\bf q}_{z}^{TE}>0$ and ${\bf q}_x^{TE}\cdot{\bf
S}_{T}^{TE}<0$, the refraction of Poynting vector is always
negative, even if the refraction of wave vector is always positive.
For TM polarized waves, $\varepsilon_z \mu_y$ is positive, if
$\varepsilon_z \mu_y<\varepsilon_I \mu_I$, the propagation occur in
the branch $-\pi/2<\theta_{I}<-\theta_{C}^{TM}$ and
$\theta_{C}^{TM}<\theta_{I}<\pi/2$. Here ${\bf k}_z\cdot{\bf
q}_{z}^{TM}>0$ and ${\bf q}_x^{TM}\cdot{\bf S}_{T}^{TM}<0$, the
Poynting vector exhibits negative refraction, while the wave-vector
presents positive refraction. If $\varepsilon_z \mu_y>\varepsilon_I
\mu_I$, the transmission phenomenon never occur in any incidence
angles. From the above anlyses, we can easily find that TE polarized
wave exhibits the total transmission, while TM polarized wave
presents total reflection.

To the best of our knowledge, the case III has not been discussed
previously, since AMMs fall into only the following distinct
groups: $\varepsilon_x=\varepsilon_y\neq\varepsilon_z$ and
$\mu_x=\mu_y\neq\mu_z$~\cite{Hu2002,Smith2003}. In our case,
however, there is no need for the elements satisfy such a
relation. Thus we can reveal a new kind of wave-vector surface. In
following analyses, we will pay more attention to the special
case.

For the purpose of illustration, we summarize the amphoteric
refractions and the corresponding incidence branch for TE and TM
waves in Table~\ref{ARIB}. Evidently, we can find that both TE and
TM waves exhibit the same positive or negative refraction. In
conventional anisotropic plasmas, only the tensor elements of
permittivity could be permitted negative. Hence it is impossible for
both TE and TM waves exhibit double-sheeted hyperboloid dispersion
relation~\cite{Chen1983,Allis1963}. In contrast to conventional
anisotropic plasma, nonmagnetic AMMs have also been constructed
recently~\cite{Podolskiy2005,Elser2006}. Since there is a distinct
lack of free magnetic poles in the real world. The only way we can
create a material with negative permeability is to
fabricate~\cite{Pendry1999}. In the present AMM, however, it is
generally accepted tensor elements of permittivity
$\boldsymbol{\varepsilon}$ and permeability $\boldsymbol{\mu}$ could
be negative. Hence the present AMM will exhibit more interesting
characteristics.

\begin{table}
\caption{\label{ARIB}Amphoteric refractions and incidence branch
for TE and TM waves. Note: X and Z indicate the revolution axes of
wave-vector surface coincide with x and z axes. P and N denote
positive and negative refraction, respectively.}
\begin{tabular}{|c c c|c c c|c|c|c|}\hline \hline
$\varepsilon_{x}$ & $\varepsilon_{y}$ & $\varepsilon_{z}$ &
$\mu_{x}$ & $\mu_{y}$ & $\mu_{z}$  &  TE Waves & TM Waves\\
\hline $+$ & $+$ &$-$ & $+$ & $+$ & $-$ & Z N $[-\pi/2, \pi/2]$ &
Z N $[-\pi/2, \pi/2]$\\
\hline$-$ & $-$ & $+$& $-$ & $-$ & $+$&
Z P $[-\pi/2, \pi/2]$ & Z P $[-\pi/2, \pi/2]$\\
\hline $-$ & $+$ & $+$& $-$ & $+$ & $+$& X P
$[-\pi/2,-\theta_{C}^{TE}] \cup [\theta_{C}^{TE}, \pi/2]$ &
X P $[-\pi/2,-\theta_{C}^{TM}] \cup [\theta_{C}^{TM}, \pi/2]$\\
\hline $+$ & $-$ & $-$& $+$ & $-$ & $-$& X N
$[-\pi/2,-\theta_{C}^{TE}] \cup [\theta_{C}^{TE}, \pi/2]$&
X N $[-\pi/2,-\theta_{C}^{TM}] \cup [\theta_{C}^{TM}, \pi/2]$\\
\hline $+$ & $+$ & $-$& $+$ & $-$ & $-$& Z N $[-\pi/2, \pi/2]$ &
X N $[-\pi/2, -\theta_{C}^{TM}] \cup [\theta_{C}^{TM}, \pi/2]$\\
\hline$-$ & $-$ & $+$& $-$ & $+$ & $+$& Z P $[-\pi/2, \pi/2]$ &
X P $[-\pi/2, -\theta_{C}^{TM}] \cup [\theta_{C}^{TM}, \pi/2]$\\
\hline $-$ & $+$ & $+$& $-$ & $-$ & $+$& X P
$[-\pi/2,-\theta_{C}^{TE}] \cup [\theta_{C}^{TE}, \pi/2]$ &
Z P $[-\pi/2, \pi/2]$\\
\hline$+$ & $-$ & $-$& $+$ & $+$ & $-$& X N
$[-\pi/2,-\theta_{C}^{TE}] \cup [\theta_{C}^{TE}, \pi/2]$  &
Z N $[-\pi/2, \pi/2]$\\
\hline \hline
\end{tabular}
\end{table}

\begin{figure}
\includegraphics[width=8cm]{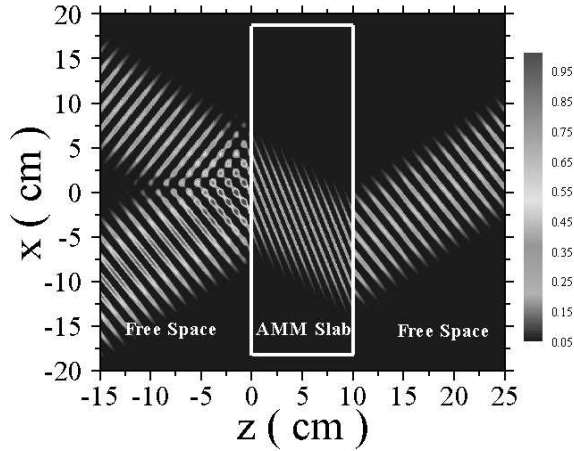}
\caption{\label{Fig4} The characteristics polarization-sensitive
propagation in an AMM slab. We choose the AMM with combination of
$\boldsymbol{\varepsilon}=[2,1,-1]$ and
$\boldsymbol{\mu}=[2,-1,-1]$. The isotropic medium is assumed as
vacuum with $\varepsilon_I=1$ and $\mu_I=1$. TE polarized beam is
totally transmitted, while TM polarized beam is totally reflected.}
\end{figure}

Now we want to enquire: what new application can be identified to
utilize the polarization-sensitive wave characterizes. Here we
introduce an idea to construct an efficient polarizing beam splitter
by using the AMM in the case III. To obtain a better picture of the
effect of beam splitter, we consider a modulated beam of finite
width. It should be pointed out that the modulated Gaussian beam has
been extensively applied to investigated negative
refraction~\cite{Kong2002,Smith2002}. Following the method outlined
by Lu~\emph{et al}~\cite{Lu2004}, let us consider a modulate
Gaussian beam with with squared magnitudes of TE and TM polarization
incident from free space into the AMM slab. A general incident wave
vector is written as ${\bf k}={\bf k}_0+{\bf k}_\perp $, where ${\bf
k}_\perp$ is perpendicular to ${\bf k}_0$ and $\omega_0=c k_0$. We
assume its Gaussian weight is
\begin{equation}
\tilde{E} (k_\perp) = \frac{w_0}{\sqrt{\pi}} \exp [- w_0^2
k_\perp^2 ],\label{fk}
\end{equation}
where $w_0$ is the spatial extent of the incident beam. We want the
Gaussian beam to be aligned with the incident direction defined by
the vector ${\bf k}_0=k_0 \cos \theta_I {\bf e}_x+ k_0 \sin \theta_I
{\bf e}_z$. For the purpose of illustration, the spatial maps of the
electric fields are plotted in Fig.~\ref{Fig4}. We set the incidence
angle equal to the Brewster angle $\theta_B^{TE}\simeq 35^\circ$,
$w_0=2$, and  $k_0=5$. We can easily find that TM polarized beam is
totally transmitted, while TE polarized beam is totally reflected.
The interesting characteristics allow us to construct a polarizing
beam splitter.

Strictly speaking, the modulated Gaussian beam we used is not
monochromatic. For a fundamental Gaussian beam, however, is
monochromatic. From the point of Fourier optics, the monochromatic
beam is considered to be composed of a series uniform plane waves
travelling in sightly different directions. Therefore the beam
splitting properties is still valid, since the monochromatic
Gaussian only modulated in wave-vector space~\cite{Goodman1996}.

It should be mentioned that the amplification of evanescent waves
could be achieved in the form of an AMM slab~\cite{Hu2002,Luo2007a}.
Hence we can expect that both TE and TM evanescent waves can be
amplified by the AMM slab. Here we want to concentrate our attention
on the polarization-sensitive effect of propagating waves. Hence we
do not involve in a detail discussion on evanescent waves. In
addition, the backward wave propagation can occur in this type of
AMM. To investigate the intriguing phenomena, we can calculate the
direction of the Poynting vector with respect to the wave vector.
The wave with ${\bf q}\cdot{\bf S}_{T}<0$ has been called the
backward wave or left-handed
wave~\cite{Lindell2001,Belov2003,Woodley2006}.

From the above anlyses, we know that the AMM exhibit an intriguing
polarization-sensitive propagation. Now we want to inquire: whether
there is a kind of AMM in which TE and TM polarized waves propagate
in the same direction. To investigate this question, we can exam the
transmission of wave-vector and Poynting vector. It is interested to
note that if the tensor elements satisfied the following condition:
\begin{equation}
\frac{\varepsilon_x}{\mu_x}=\frac{\varepsilon_y}{
\mu_y}=\frac{\varepsilon_z}{\mu_z}=C
\end{equation}
where $C$ is a constant. If $C>0$, the AMM will exhibit a
polarization-insensitive propagation. In this case, both TE and TM
polarized waves present the same wave-vector surface, such as a
ellipsoid or a double-sheeted hyperboloid. The two polarized waves
will exhibit the same propagating characteristic, and this AMM can
be regard as quasi-isotropic~\cite{Luo2006a}. If $C<0$, TE and TM
polarized waves also present the same single-sheeted hyperboloid
wave-vector surface. However the two polarized waves will exhibit
opposite amphoteric refractions~\cite{Luo2007c}.

From the table I, we have found that $\varepsilon_z$ and $\mu_z$
always exhibit the same sign. Hence TE and TM waves will present the
same positive or negative refraction. It is generally believed that
the two polarized waves can not exhibit the opposite amphoteric
refractions. Now a question naturally arise: how can the opposite
amphoteric refractions can be realized in the AMM? In the following
section, we want to explore this intriguing problem in detail.

\section{Polarization-sensitive propagation}
\begin{figure}
\includegraphics[width=6cm]{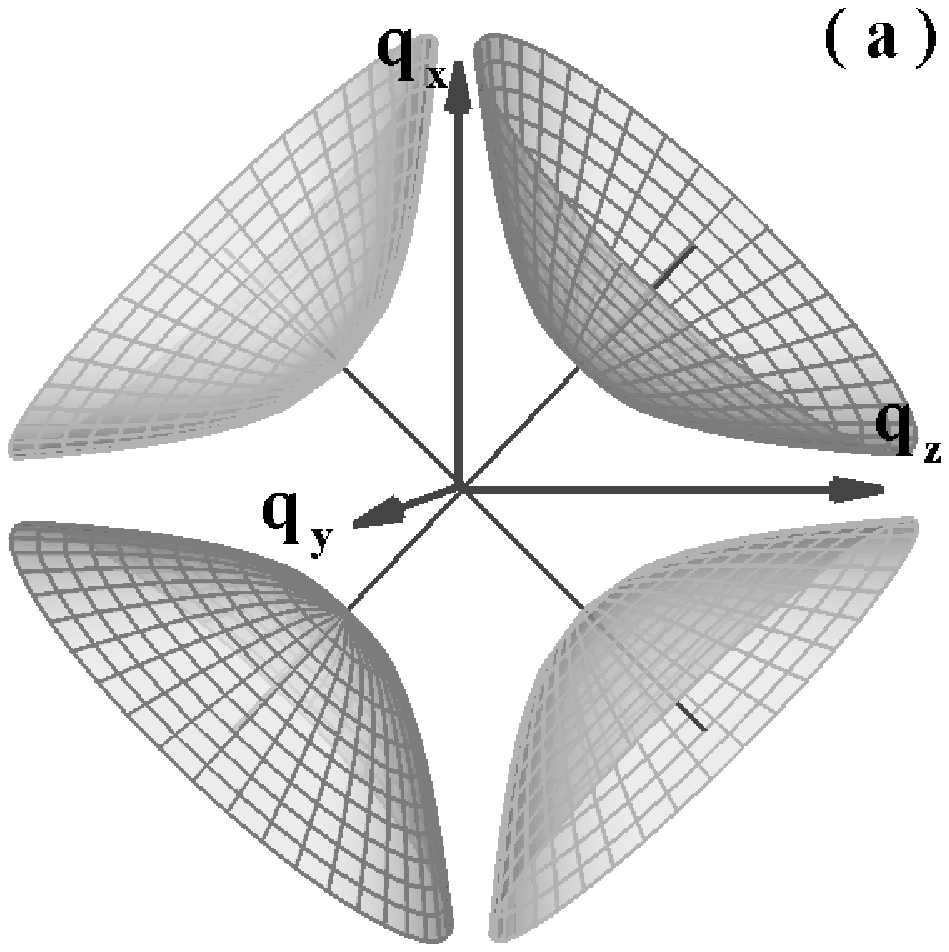}~~~~~~~~
\includegraphics[width=6cm]{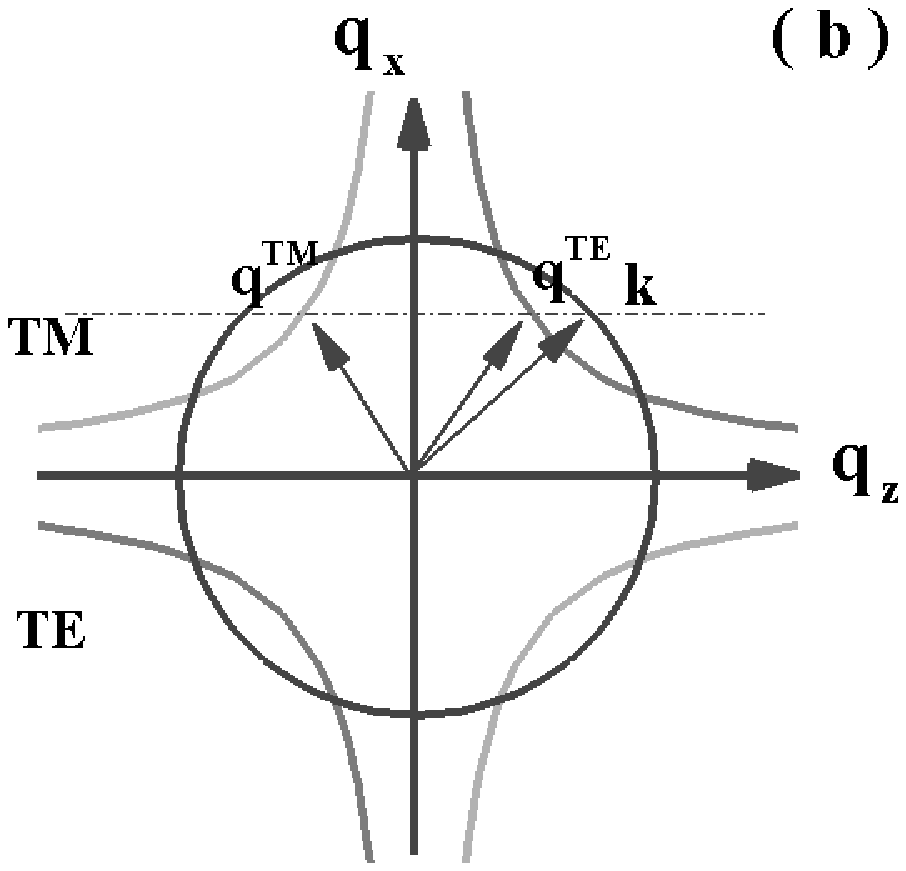}
\caption{\label{Fig5} We assume there is an angle $\varphi$ between
the principle axis and the propagation axis. (a) The 3D frequency
contours present propagation characteristics of TE (black) and TM
(gray) waves. (b) The circle and the double-sheeted hyperbolas
represent the frequency contours of isotropic media and AMM,
respectively. }
\end{figure}

The AMMs with rotating principle axis can exhibit some interesting
physics phenomenons, such as anomalous negative
refraction~\cite{Zhang2003,Luo2005,Depine2006}, superluminal or
subluminal group propagation~\cite{Luo2006b}, and large beam
shift~\cite{Zhong2006,Wang2006}. Here we want to discuss the
anomalous amphoteric refractions. We assume that there is an angle
$\varphi$ between the principle axis and the propagation axis. For
TE polarized wave, the Maxwell's equations yield the dispersion
relation in the AMM as
\begin{equation}
\alpha q_{x}^2+\beta q_{z}^2+\gamma
q_{x}q_{z}=\frac{\omega^2}{c^2}. \label{D2}
\end{equation}
Here $q_{x}$ and $q_{z}$ represent the $x$ and $z$ components of
transmitted wave vector in the propagating coordinate system. The
parameters $\alpha$, $\beta$ and $\gamma$ are given by
\begin{eqnarray}
\alpha &=&\frac{1}{\varepsilon_y \mu_x \mu_z}
(\mu_x \cos^2\varphi+\mu_z \sin^2\varphi),\nonumber\\
\beta &=&\frac{1}{ \varepsilon_y \mu_x \mu_z}
(\mu_x \sin^2\varphi+\mu_z \cos^2\varphi),\nonumber\\
\gamma &=&\frac{1}{\varepsilon_y \mu_x \mu_z} (\mu_z \sin 2\varphi
-\mu_x \sin 2\varphi).
\end{eqnarray}
The corresponding 3D dispersion geometry is shown in
Fig.~\ref{Fig5}(a). We can find that it is a rotating manipulation
of case III. The refraction diagram in  $x-z$  plane is plotted in
Fig.~\ref{Fig5}(b), where a plane electromagnetic wave is incident
from free space into the AMM.

The $z$-component of the wave vector can be found by the solution
of Eq.~(\ref{D2}), which yields
\begin{equation}
 q_z^{TE} = \frac{1}{2 \beta}\bigg[\sigma\sqrt {4\beta \frac{\omega^2}{c^2}+(\gamma^2-4
\alpha \beta )q_x^2}-\gamma q_x\bigg], \label{qz}
\end{equation}
Here $\sigma=\pm1$, the choice of sign ensures that light power
propagates away from the surface to the $+z$ direction. The values
of refraction wave-vector can be found by using the boundary
condition and hyperbolic dispersion relation.

Now a question easily be asked: how to determine the positive or
negative refraction in the special case? To distinguish the
positive and negative refraction, we can calculate the direction
of the Poynting vector with respect to the wave vector:
\begin{equation}
{\bf q}_x^{TE} \cdot {\bf S}_{T}^{TE}=\frac{4\varepsilon_y k_x
k_z^2 (2 \alpha k_x+ \gamma q_z^{TE}) E_0^2}{(2 k_z + 2
\varepsilon_y \beta q_z^{TE}+\varepsilon_y \gamma
k_x)^2}.\label{IC}
\end{equation}
For TM polarized wave, $S_{T}^{TM}$ can be obtained by exchanging
$\varepsilon_i$ and $\mu_i$. Hence we can determine that TE and TM
polarized waves exhibit the opposite amphoteric refractions in such
an AMM, such that TE polarized wave is positively refracted whereas
TM polarized wave is negatively refracted. The opposite amphoteric
refractions will result in a large birefringence.

\begin{figure}
\includegraphics[width=8cm]{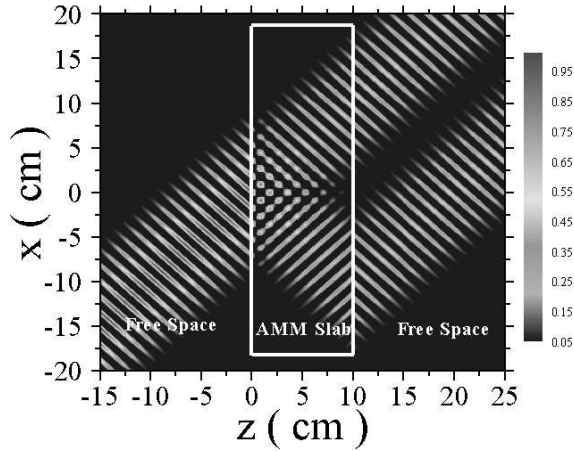}
\caption{\label{Fig6} The characteristics polarization-sensitive
propagation in an AMM slab. We choose the AMM with combination of
$\boldsymbol{\varepsilon}=[1,1,-1]$ and
$\boldsymbol{\mu}=[1,-1,-1]$. The isotropic medium is assumed as
vacuum with $\varepsilon_I=1$ and $\mu_I=1$. TM polarized beam is
negatively refracted, while TE polarized beam is positively
refracted.}
\end{figure}

Finally, we want to introduce another type of polarizing beam
splitter, which is based on the opposite amphoteric refractions. For
the purpose of illustration, the spatial maps of the electric fields
are plotted in Fig.~\ref{Fig6}. We can choose the appropriate medium
parameters, then the reflections of TE and TM polarized waves are
completely absent. A large beam splitting angle between the two
polarized waves can be obtained as $ 90^\circ$. Compared with the
polarizing beam splitters made from  the conventional anisotropic
crystal, the present counterpart is more simple and more efficient.
However the important limitation in practical realization of the
splitter is loss in the AMM slab. We trust that it is advantageous
to employ ultralow-loss AMMs to construct a very simple and very
efficient splitter~\cite{Luo2007b}.

In the above analyses, the interesting effects of
polarization-sensitive propagation in AMM are discussed using the 3D
frequency contours. From a purely wave propagation point of view,
the values of permittivity and permeability tensor elements were
taken as constants at a fixed frequency. A question can then be
asked in regard to the influence of frequency dispersion on the
above analysis. In principle, the geometry of the wave-vector
surface is determined by the signs of the medium parameters. In a
certain frequency region where the permittivity and permeability
tensor elements change signs, the corresponding wave-vector surfaces
will present and a new characteristics of wave propagation emerges.
Hence we expect our analyses can be extended to study the general
behavior of other possible combinations.

\section{Conclusion }\label{sec4}
In conclusion, we have investigated polarization-sensitive
propagation in the AMM with double-sheeted hyperboloid dispersion
relation. Under appropriate conditions, the anomalous
omnidirectional total reflection and oblique total refraction can
occur in the interface associated with the AMM. We are especially
interested in the case that one polarized wave is totally refracted,
the other is totally reflected. We have studied the opposite
amphoteric refractions, such that one polarized wave exhibits
positive refraction, while the other presents negative refraction.
Based on the polarizing-sensitive propagation, we have introduced
two types of polarizing beam splitters. We are sure that our scheme
has not exhausted the interesting properties. The wave
characteristics of polarization-sensitive propagation could be taken
advantage of to other practical applications, such as polarizers,
beam filters, and polarization dependent lenses.

\begin{acknowledgements}
This work was supported by projects of the National Natural
Science Foundation of China (No. 10125521  and  No. 10535010), the
973 National Major State Basic Research and Development of China
(No. G2000077400), and Major State Basic Research Developing
Program (No. 2007CB815000).
\end{acknowledgements}

\end{document}